# Description of non-specific DNA-protein interaction and facilitated diffusion with a dynamical model

Ana-Maria FLORESCU and Marc JOYEUX [(#)]

*Laboratoire de Spectrométrie Physique (CNRS UMR 5588),*

*Université Joseph Fourier - Grenoble 1,*

*BP 87, 38402 St Martin d'Hères, FRANCE*

**Abstract** : We propose a dynamical model for non-specific DNA-protein interaction, which is based on the "bead-spring" model previously developed by other groups, and investigate its properties using Brownian Dynamics simulations. We show that the model successfully reproduces some of the observed properties of real systems and predictions of kinetic models. For example, sampling of the DNA sequence by the protein proceeds via a succession of 3d motion in the solvent, 1d sliding along the sequence, short hops between neighboring sites, and intersegmental transfers. Moreover, facilitated diffusion takes place in a certain range of values of the protein effective charge, that is, the combination of 1d sliding and 3d motion leads to faster DNA sampling than pure 3d motion. At last, the number of base pairs visited during a sliding event is comparable to the values deduced from single-molecule experiments. We also point out and discuss some discrepancies between the predictions of this model and some recent experimental results as well as some hypotheses and predictions of kinetic models.

[(#)] email : marc.joyeux@spectro.ujf-grenoble.fr

# 1 - INTRODUCTION

Gene expression is regulated in both prokaryotes and eukaryotes by proteins called transcription factors, which bind to chromosomal DNA at specific sites and subsequently promote or prevent the transcription by RNA polymerase. It is believed that the targeting of transcription factors to their sites is a purely passive process, in the sense that these proteins simply wander through the nucleus or the cell till they find their sites. Almost four decades ago, Riggs, Bourgeois and Cohn pointed out that the LacI transcription factor of Escherichia coli finds its binding site hundred times faster than expected by three-dimensional (3d) diffusion in the solvent [1]. Although this initial work on the LacI repressor has been reinterpreted to some extent [2-6], recent experiments using a rapid footprinting procedure that reveals the occupancy of target sites confirm the initially proposed value for the rate constant [7]. Admittedly, it now appears that the LacI repressor is an exception rather than the rule. Indeed, while a few other systems were found to display rates larger than the 3d diffusion limit [8,9], most of the investigated DNA binding proteins have association rate constants that are close to this limit [10-16]. Still, the initial claim of Riggs et al triggered a lot of theoretical work, which was essentially aimed at understanding how such large rates are possible, but also shed more general light on the mechanism of non-specific DNA-protein interactions (see for example refs. [17-23] and references therein). As a matter of fact, it is now generally admitted that proteins alternate between 3d motion through the volume of the cell and one-dimensional (1d) diffusion (sliding) along the DNA, a process which is usually called "facilitated diffusion". Most importantly, these theoretical speculations have been confirmed by single molecule experiments, in which individual proteins sliding along DNA could be visualized [24-29]. In order to complete the description of site targeting by proteins, *specific* DNA-protein interactions, that is, the description of how proteins recognize their specific binding sites during 1d sliding sequences [30], are a current hot topic of theoretical biophysics [21,31-35].

The point is that essentially all the models that have been developed up to now to mimic DNA-protein non-specific interactions (facilitated diffusion) are mass action kinetic models with phenomenological rates [17-23], which are based on an *a priori* scenario for site targeting. More precisely, these models *assume* that proteins alternate between 3d motion through the volume of the cell and 1d sliding along the DNA and that both 1d and 3d motions are random walks characterized by diffusion coefficients $D_{1d}$ and $D_{3d}$. These basic assumptions are then used to estimate the expressions of various quantities of interest, like the association rate of binding and the total time required to find the target, as a function of a set of well-defined geometric quantities, like the sequence length $L$ and the average volume $V$ occupied by one site, and a more or less extended list of rate constants and reactions probabilities (see for example Table I of [17]). For example, Halford and Marko obtained that the reaction rate (i.e. the inverse of the total search time) for unit protein concentration may be expressed as [20]

$$k = \left( \frac{1}{D_{3d}\, \ell_{\text{sl}}} + \frac{L\, \ell_{\text{sl}}}{D_{1d}\, V} \right)^{-1} , \qquad (1.1)$$

where $\ell_{\text{sl}}$, the characteristic sliding length, is proportional to the inverse of the probability of dissociation of the protein per unit sliding length. Since the 3d diffusion-limited rate is just $a\, D_{3d}$, where $a$ is the target site size, it follows that the acceleration of the reaction due to facilitated diffusion is $k/(a\, D_{3d})$. After a couple of additional hypotheses, Halford and Marko concluded that this ratio is at maximum equal to about 30 for an optimal sliding length $\ell_{\text{sl}} \approx 100$ base pairs, close to values obtained from single-molecule experiments [26-28].

Although they do provide invaluable information in many areas of biophysics, and especially in the field of protein-DNA interactions, kinetic models therefore have the limitation that they do not really indicate how site targeting takes place, in the sense that the mechanism of site targeting is supposed to be known *a priori*. Most of the hypotheses underlying kinetic models are firmly grounded and supported by experimental evidence. For example, there is nowadays little doubt that site targeting proceeds through an alternation of 3d motion through the volume of the cell and 1d sliding along the DNA. It still remains that some of the hypotheses are more questionable. For example, the density of DNA in the cell is so large that one might wonder whether 3d motion is indeed best described as a purely diffusive process. Moreover, kinetic models neglect any contribution from electrostatic interactions between DNA and the protein, while such interactions necessarily exist and may lead to association rate constants that exceed by far the 3d diffusion limit $a D_{3d}$ [20].

It therefore appears as a necessity to backup and confirm the assumptions of kinetic models and the conclusions derived there from thanks to calculations based on completely different models. The purpose of the present paper is precisely to propose a dynamical model (i.e. a molecular mechanical model) for non-specific DNA-protein interactions and check to what extent results obtained with this model match the hypotheses and conclusions of kinetic ones. By "dynamical" or "molecular mechanical" model, we mean a microscopic model which relies uniquely on the definition of a Hamiltonian describing all possible interactions inside the model cell (i.e. the interactions between DNA and DNA, DNA and protein, and DNA/protein and cell wall) and the choice of equations of the motion, which should take into account as realistically as possible the effect of the solvent. For this purpose, we adapted the "wormlike-chain and beads" or "bead-spring" model developed by several groups [36-41], paying special attention to the term describing the interaction between DNA and the protein, and propagated trajectories using the Brownian Dynamics algorithm of Ermak and McCammon [42], which includes hydrodynamic interactions.

It is emphasized that comparison between the two types of models, kinetic and dynamical ones, is all the more meaningful, as it is generally not possible to draw a direct correspondence between the statistical quantities involved in statistical models and the microscopic quantities on which dynamical models rely. For example, the characteristic sliding length $\ell_{sl}$, which appears in Eq. (1.1), certainly depends on (i) the depth, width and shape of the attractive DNA-protein interaction term, (ii) the hydrodynamic radii of DNA and the protein, (iii) temperature, but this dependence is quite complex and we do not know of an expression that would relate these quantities. Moreover, dynamical models can shed additional light compared to kinetic ones with respect to a certain number of points. For instance, the effect of temperature variation, e.g. on the ratio of 1d sliding and 3d motion, can easily be predicted from dynamical simulations, while this is basically an input quantity of kinetic models.

The remainder of this article is organized as follows. The model we propose is described in Sect. 2 along with the Hamiltonian which governs the interactions between its various components (DNA, protein and cell wall) and the equations of motions that were used to integrate trajectories. Results obtained with this model are presented in Sect. 3 and compared to the assumptions and results of statistical models. Finally, we discuss in Sect. 4 the validity of the model and how it could be improved.

## 2 – MODEL AND SIMULATIONS

The model system consists of a cell, which contains a protein and several DNA segments. The cell is taken as a sphere of radius $R_0$, while the protein is modelled as a single bead with hydrodynamic radius $a_{PROT} = 3.5$ nm and an effective charge $e_{PROT}$ placed at its centre [40,41]. As in the work of Schlick and co-workers [36], each DNA segment consists of a chain of $n$ beads separated at equilibrium by a distance $l_0 = 5.0$ nm. Each bead represents 15 base pairs, has a hydrodynamic radius $a_{DNA} = 1.78$ nm, and an effective charge $e_{DNA} = 0.243 \times 10^{10} \, l_0 \, \overline{e} \approx 12 \, \overline{e}$ placed at its centre ($\overline{e}$ is the charge of the electron). $n$ is chosen so that the length of each DNA segment is approximately equal to the radius of the cell, i.e. $n l_0 \approx R_0$, in order that the cell is more or less homogeneously filled with DNA but excessive curvature of DNA segments touching the cell wall is avoided. The number $m$ of segments is chosen so that the density of bases inside the cell is close to the experimentally observed one. As pointed out in [20], the volume $V$ of the cell is connected to the total DNA length $L$ according to $V = w^2 L$, where $w$ represents roughly the spacing of nearby DNA segments. $m$ must therefore fulfil the relation $\frac{4}{3}\pi R_0^3 \approx w^2 m n l_0$, where the average value $w = 45.0$ nm holds for both prokaryote and eukaryote cells. Practically, we essentially worked with a sequence consisting of $m = 50$ segments of $n = 40$ beads (i.e. a total of 30000 base pairs) and a cell radius $R_0 = 0.169$ μm, but we also performed calculations for a smaller system ($m = 30$, $n = 33$, $R_0 = 0.134$ μm) and a larger one ($m = 80$, $n = 50$, $R_0 = 0.213$ μm) to check the effect of the sequence length on the obtained results.

The potential energy $E_{pot}$ of the system consists of three terms

$$E_{pot} = V_{DNA} + V_{DNA/PROT} + V_{wall} \ , \tag{2.1}$$

where $V_{DNA}$ describes the potential energy of the DNA segments and the interactions between them, $V_{DNA/PROT}$ stands for the interactions between the protein bead and DNA segments, and $V_{wall}$ models the interactions with the cell wall, which maintain the protein bead and the DNA segments inside the cell. $V_{DNA}$ is borrowed from Schlick and co-workers [36] :

$$V_{\text{DNA}} = E_{\text{s}} + E_{\text{b}} + E_{\text{e}}$$

$$E_{\text{s}} = \frac{h}{2} \sum_{j=1}^{m} \sum_{k=1}^{n-1} \left( l_{j,k} - l_0 \right)^2$$

$$E_{\text{b}} = \frac{g}{2} \sum_{j=1}^{m} \sum_{k=1}^{n-2} \theta_{j,k}^2$$

$$E_{\text{e}} = \frac{e_{\text{DNA}}^2}{4\pi\varepsilon} \sum_{j=1}^{m} \sum_{k=1}^{n-2} \sum_{K=k+2}^{n} \frac{\exp\left( -\frac{1}{r_D} \left\| \mathbf{r}_{j,k} - \mathbf{r}_{j,K} \right\| \right)}{\left\| \mathbf{r}_{j,k} - \mathbf{r}_{j,K} \right\|}$$

$$+ \frac{e_{\text{DNA}}^2}{4\pi\varepsilon} \sum_{j=1}^{m} \sum_{k=1}^{n} \sum_{J=j+1}^{m} \sum_{K=1}^{n} \frac{\exp\left( -\frac{1}{r_D} \left\| \mathbf{r}_{j,k} - \mathbf{r}_{J,K} \right\| \right)}{\left\| \mathbf{r}_{j,k} - \mathbf{r}_{J,K} \right\|} \quad , \tag{2.2}$$

where $\mathbf{r}_{j,k}$ denotes the position of bead $k$ of segment $j$, $l_{j,k} = \left\| \mathbf{r}_{j,k} - \mathbf{r}_{j,k+1} \right\|$ the distance between two successive beads belonging to the same segment, and $\theta_{j,k}$ the angle formed by three successive beads on the same segment

$$\cos\theta_{j,k} = \frac{\left( \mathbf{r}_{j,k} - \mathbf{r}_{j,k+1} \right) \cdot \left( \mathbf{r}_{j,k+1} - \mathbf{r}_{j,k+2} \right)}{\left\| \mathbf{r}_{j,k} - \mathbf{r}_{j,k+1} \right\| \left\| \mathbf{r}_{j,k+1} - \mathbf{r}_{j,k+2} \right\|} . \tag{2.3}$$

$E_{\text{s}}$ is the bond stretching energy. This is actually a computational device without real biological meaning, which is essentially aimed at avoiding having to deal with rigid rods. The stretching force constant is fixed at $h = 100\, k_{\text{B}}T / l_0^2$, with $T = 298$ K (see the discussion in [36] for this choice for $h$). $E_{\text{b}}$ is the elastic bending potential. The bending rigidity constant, $g = 9.82\, k_{\text{B}}T$, is fixed so as to provide the correct persistence length $p = 50.0$ nm (i.e. 10 beads) [36,43]. $E_{\text{e}}$ is a Debye-Hückel potential which describes repulsive electrostatic interactions between DNA beads [36,44,45]. In eq. (2.2), $r_D = 3.07$ nm stands for the Debye length at 0.01 M molar salt concentration of monovalent ions [36] and $\varepsilon = 80\,\varepsilon_0$ for the dielectric constant of the solvent. Note that electrostatic interactions between neighbouring beads belonging to the same segment are not included in the expression of $E_{\text{e}}$ in eq. (2.2), because it is considered that these nearest-neighbour interactions rather contribute to the stretching and bending terms.

The potential $V_{\text{wall}}$, which models the interactions between DNA and the protein and the cell wall, is taken as a sum of repulsive terms that act on the beads that trespass the radius of the cell, $R_0$, and repel them back inside the cell

$$V_{\text{wall}} = k_B T \sum_{j=1}^{m} \sum_{k=1}^{n} f\left(\|\mathbf{r}_{j,k}\|\right) + 10\, k_B T\, f\left(\|\mathbf{r}_{\text{PROT}}\|\right) \tag{2.4}$$

where $\mathbf{r}_{\text{PROT}}$ denotes the position of the protein and $f$ is a function defined as

if $x \leq R_0$ : $f(x) = 0$

if $x > R_0$ : $f(x) = \left(\dfrac{x}{R_0}\right)^6 - 1$ . $\tag{2.5}$

The coefficients $k_B T$ and $10\, k_B T$ in Eq. (2.4) were roughly adjusted by hand, in order that, at 298 K and for cell radii $R_0$ comprised between 0.134 and 0.213 µm, all the beads (DNA and protein) remain confined inside a sphere of radius $\approx 1.10\, R_0$, which insures that the time spent by the beads outside the cell is negligible. The coefficient is 10 times larger for the protein bead than for the DNA ones, because the protein is modelled by a single bead, so that its mobility is much larger than that of the interconnected DNA beads and its motion outside the sphere of radius $R_0$ more difficult to oppose.

Last but not least, the interaction $V_{\text{DNA/PROT}}$ between the protein and DNA beads is the sum of an attractive and a repulsive term

$$V_{\text{DNA/PROT}} = E_{\text{e}}^{(\text{P})} + E_{\text{ev}}$$

$$E_{\text{e}}^{(\text{P})} = -\frac{e_{\text{DNA}}\, e_{\text{PROT}}}{4\pi\varepsilon} \sum_{j=1}^{m} \sum_{k=1}^{n} \frac{\exp\left(-\dfrac{1}{r_D}\|\mathbf{r}_{j,k} - \mathbf{r}_{\text{PROT}}\|\right)}{\|\mathbf{r}_{j,k} - \mathbf{r}_{\text{PROT}}\|} \tag{2.6}$$

$$E_{\text{ev}} = k_B T\, \frac{e_{\text{PROT}}}{e_{\text{DNA}}} \sum_{j=1}^{m} \sum_{k=1}^{n} F\left(\|\mathbf{r}_{j,k} - \mathbf{r}_{\text{PROT}}\|\right)$$

where $F$ is a function defined as

$$\text{if } x \leq \sqrt{2}\,\sigma \,:\, F(x) = 4\left(\left(\frac{\sigma}{x}\right)^4 - \left(\frac{\sigma}{x}\right)^2\right) + 1$$

$$\text{if } x > \sqrt{2}\,\sigma \,:\, F(x) = 0 \tag{2.7}$$

and $\sigma = a_{\text{DNA}} + a_{\text{PROT}} = 5.28$ nm. $E_{\text{e}}^{(\text{P})}$ is the Debye-Hückel potential, which models the attractive electrostatic interactions between the protein and DNA beads, while $E_{\text{ev}}$ is an excluded volume term, which prevents the protein bead from sticking to a DNA bead and $E_{\text{e}}^{(\text{P})}$ from diverging. $E_{\text{ev}}$ is sometimes taken as the repulsive part of the Lennard-Jones potential [39]. Being of order 12, this function is however so sharp that it leads too often to numerical bugs, while the order 4 function $F(x)$ enables trouble-free calculations. The prefactor of $E_{\text{ev}}$ was chosen as $k_{\text{B}}T\, e_{\text{PROT}}\,/\,e_{\text{DNA}}$, because this insures that the DNA/protein interaction $V_{\text{DNA/PROT}}$ displays a global minimum very close to $\sigma = a_{\text{DNA}} + a_{\text{PROT}}$, whatever the charge $e_{\text{PROT}}$ of the protein bead (see Fig. 1). Intuitively, $V_{\text{DNA/PROT}}$ must indeed be minimum at some value close to the sum of the radii of DNA and the protein (which is close to $\sigma$) in order for 1d sliding to take place. Moreover, we will take advantage of the fact that the position of this minimum does not depend on $e_{\text{PROT}}$ to let $e_{\text{PROT}}$ assume different values, thereby varying the percentage of time the protein bead spends in 1d sliding and 3d motion (see below).

The Brownian Dynamics algorithm of Ermak and McCammon [42] is based on a simplification of the generalized Langevin equations, which holds for small inertial contributions and sufficiently large time steps. According to this first-order algorithm, the updated position vector for the beads, $\mathbf{r}^{(n+1)}$, is obtained from the current position vector, $\mathbf{r}^{(n)}$, according to

$$\mathbf{r}^{(n+1)} = \mathbf{r}^{(n)} + \frac{\Delta t}{k_{\text{B}}T}\mathbf{D}^{(n)}.\mathbf{F}^{(n)} + \sqrt{2\Delta t}\,\mathbf{L}^{(n)}.\zeta^{(n)} \,, \tag{2.8}$$

where $\Delta t$ is the time step. Note that $\mathbf{r}^{(n)}$ and $\mathbf{r}^{(n+1)}$ are collective vectors that include the position vectors $\mathbf{r}_{j,k}$ of all DNA beads, as well as the position vector $\mathbf{r}_{\mathrm{PROT}}$ of the protein bead, at steps $n$ and $n+1$. The second term in the right-hand side of eq. (2.8) models the diffusive effects of the solvent. $\mathbf{F}^{(n)}$ is the collective vector of inter-particle forces arising from the potential energy $E_{\mathrm{pot}}$ and $\mathbf{D}^{(n)}$ the hydrodynamic interaction diffusion tensor. As in [39], we built the successive tensors $\mathbf{D}^{(n)}$ using a modified form of the Rotne-Prager tensor for unequal size beads [46-48] (see eqs. (26)-(28) of [39]). The third term in the right-hand side of eq. (2.8) models the effects on $\mathbf{r}^{(n+1)}$ of collisions between the solvent and the protein and DNA beads. $\xi^{(n)}$ is a vector of random numbers extracted at each step $n$ from a Gaussian distribution of mean 0 and variance 1 and $\mathbf{L}^{(n)}$ is the lower triangular matrix obtained from the Choleski factorization of $\mathbf{D}^{(n)}$

$$\mathbf{D}^{(n)} = \mathbf{L}^{(n)} \cdot {}^{t}\mathbf{L}^{(n)} \tag{2.9}$$

where ${}^{t}\mathbf{L}^{(n)}$ denotes the transpose of $\mathbf{L}^{(n)}$. The CPU time required to factor the diffusion matrix increases as the cube of the number of beads that are taken into account in $\mathbf{D}^{(n)}$, so that the Choleski factorization of $\mathbf{D}^{(n)}$ turns out to be the limiting step for the investigation of the dynamics of large systems. Fixman's approximation can be used to decrease the exponent from 3 to 2.25 [49,50], but we chose to use a more drastic approximation. Indeed, in this work we are only interested in the interaction between DNA and the protein, so that it is important that the motion of DNA close to the protein be modelled correctly. In contrast, results are little affected if the motion of DNA far from the protein is handled in a cruder way. Therefore, we used eqs. (2.8)-(2.9) to calculate the position at each time step of the protein and the 100 DNA beads closest to it, while the positions of the remaining DNA beads were obtained from the diagonal approximation of eq. (2.8), that is

$$\mathbf{r}^{(n+1)} = \mathbf{r}^{(n)} + \frac{\Delta t}{6\pi\eta \ a_{\mathrm{DNA}}} \mathbf{F}^{(n)} + \sqrt{\frac{2 \ k_B T \ \Delta t}{6\pi\eta \ a_{\mathrm{DNA}}}} \ \xi^{(n)} , \qquad (2.10)$$

where $\eta = 0.00089$ Pa s denotes the viscosity of the solvent at 298 K. Note that eq. (2.10) is just the first-order discretization of the usual Langevin equation without hydrodynamic interactions and with the second-order term arising from kinetic energy dropped. When considering a system with 2000 DNA beads, use of eqs. (2.8)-(2.9) to update the positions of the protein and the 100 closest DNA beads slows down calculations by only 10% compared to the case where eq. (2.10) is used for all beads. In contrast, the CPU time is already multiplied by a factor larger than 2 if eqs. (2.8)-(2.9) are used for the 200 DNA beads closest to the protein. On the other hand, we checked that use of eq. (2.10) to update the position of all beads leads to results that differ substantially from those presented in the remainder of this paper, while use of eqs. (2.8)-(2.9) to update the position of the 200 DNA beads closest to the protein, instead of the 100 closest ones, leads to similar results. Use of eqs. (2.8)-(2.9) for the 100 DNA beads closest to the protein therefore appears as a very reasonable choice.

For all simulations, the $m$ DNA segments were first placed inside the cell according to a randomization procedure that insures an essentially uniform distribution of the beads in the cell (see Fig. 2). The protein bead was then placed at random in a sphere of radius $R_0 / 5$. In order to avoid too strong repelling interactions at time $t = 0$, all initial configurations where the distance between the protein and at least one DNA bead turned out to be smaller than $\sigma = a_{\mathrm{DNA}} + a_{\mathrm{PROT}} = 5.28$ nm were however rejected. The equations of motion (2.8)-(2.10) were then integrated for 10 μs, in order for the system to equilibrate at the correct temperature. The quantities of interest were subsequently obtained by integrating the equations of motion for longer time intervals and averaging over several different trajectories. The Brownian Dynamics algorithm of Ermak and McCammon [42] is based on the assumption that the motions of interest occur on a time scale much longer than $M / (6\pi\eta a)$, where $M$ and $a$ are the mass and hydrodynamic radius of a bead. For the model described above, this sets a lower bound $\Delta t \gg 1$ ps. As illustrated in Fig. 3 for the time evolution of the number $N(t)$ of *different* DNA beads visited by the protein at time $t$, we accordingly checked that time steps $\Delta t$ equal to 25, 100 and 400 ps lead to identical results. Most of the results discussed below were consequently obtained with $\Delta t = 100$ ps, although a few ones dealing with the system with 4000 DNA beads were obtained with $\Delta t = 400$ ps.

## 3 – RESULTS

For the repulsive $V_{\text{DNA/PROT}}$ potential of Fig. 1, DNA and the protein never attract each other. The protein therefore moves almost freely in the solvent, except that it is repelled by the excluded volume interaction $E_{\text{ev}}$ whenever the distance to a DNA bead becomes too small. Because of the large density of DNA beads, the probability for the protein to be found close to a DNA bead is not negligible : if one considers that the protein interacts with bead $k$ of DNA segment $j$ when $\left\| \mathbf{r}_{j,k} - \mathbf{r}_{\text{PROT}} \right\| \leq \sigma$, then DNA "fills" about 3% of the cell volume and the protein is expected to spend approximately the same amount of time interacting with DNA, in spite of the absence of attractive interactions. This is indeed the case, as can be checked in Fig. 4, which shows the portion of time $\rho_{\text{1d}}$ during which the protein interacts with a DNA bead as a function of the ratio $e_{\text{PROT}} / e_{\text{DNA}}$. In this plot, the points at $e_{\text{PROT}} / e_{\text{DNA}} = 0$ precisely correspond to the repulsive potential of Fig. 1, while circles and lozenges respectively denote results obtained with the $\left\| \mathbf{r}_{j,k} - \mathbf{r}_{\text{PROT}} \right\| \leq \sigma$ and $\left\| \mathbf{r}_{j,k} - \mathbf{r}_{\text{PROT}} \right\| \leq 1.5\,\sigma$ criterions for interacting beads. It is seen that $\rho_{\text{1d}}$ is indeed close to 3% for the repulsive potential and the $\left\| \mathbf{r}_{j,k} - \mathbf{r}_{\text{PROT}} \right\| \leq \sigma$ criterion.

Because of either these not-so-infrequent collisions with DNA or the fact that our dynamical model takes electrostatic interactions into account, while kinetic models usually do not, the number $N(t)$ of different DNA beads visited by the protein in the absence of attractive terms in $V_{DNA/PROT}$ does not follow the square root law which would be expected for a purely diffusive process. This evolution, which is shown for the system with 2000 DNA beads as a dotted line in Fig. 5, can instead be modelled by a law of the form

$$\frac{N(t)}{mn} = 1 - \exp\left(-\kappa \frac{t}{mn}\right),$$ (3.1)

where $\kappa = 1.09 \ \mu s^{-1}$ (see Fig. 6). We will come back to this law shortly, but it is important to realize that it implies that $N(t)$ increases linearly at rate $\kappa$ as long as $N$ remains sufficiently small compared to the total number $mn$ of DNA beads inside the cell, while the rate steadily decreases down to zero when $N$ comes closer and closer to $mn$.

In contrast, if $e_{PROT}/e_{DNA} > 0$, then the interaction $V_{DNA/PROT}$ between the protein and DNA beads displays a minimum close to $\sigma = a_{DNA} + a_{PROT}$ (see Fig. 1), so that the motion of the protein results from the balance of conflicting constraints : $V_{DNA/PROT}$ tends to localize the protein close to DNA segments, while stochastic interactions with the solvent tend to release the protein bead in the bulk of the cell. Fig. 4 indicates that the motion of the protein therefore consists of a combination of 1d sliding and 3d motion for values of $e_{PROT}/e_{DNA}$ not too large, say, up to $e_{PROT}/e_{DNA} \approx 3$. For larger values of $e_{PROT}/e_{DNA}$, the electrostatic attraction between the protein and DNA is predominant, so that the protein spends most of the time in the neighbourhood of a DNA segment. Note, that $e_{PROT}/e_{DNA} \approx 1$ corresponds to an effective protein charge $e_{PROT} \approx 12\,\bar{e}$, which is of the same order of magnitude as experimentally determined protein effective charges [51,52].

At this point, it should be mentioned that hydrodynamic interactions tend to decrease the portion of time spent in 1d diffusion compared to 3d motion in the solvent. For example, if one neglects all hydrodynamic interactions and uses eq. (2.10) to update the position of all beads, then $\rho_{1d}$ is found to be equal to 0.60 (respectively, 0.95) for $e_{PROT}/e_{DNA} = 1$ and the $\left\| \mathbf{r}_{j,k} - \mathbf{r}_{PROT} \right\| \leq \sigma$ criterion (respectively, the $\left\| \mathbf{r}_{j,k} - \mathbf{r}_{PROT} \right\| \leq 1.5\,\sigma$ criterion) for interacting beads, instead of $\rho_{1d} = 0.20$ and 0.44. As will be discussed below, this has marked consequences on the number $N(t)$ of different beads visited by the protein at time $t$.

Fig. 7 illustrates the typical trajectory of a protein bead for the ratio $e_{PROT}/e_{DNA} = 1$. During the 15 μs time interval displayed in this figure, the protein visits four different segments. Globally, 1d sliding along each segment can last several μs, but it is frequently interrupted by shorter time intervals during which the protein is released in the solvent and at the end of which it reattaches to the same segment either at the same position or at a neighbouring one. These short jumps are often called "hops" [17,20,25,26]. On the other hand, the protein sometimes moves almost freely and for longer time intervals (several μs) in the solution before reattaching to another segment or eventually to the same segment but at a rather different position. Note also that "intersegmental transfer", which involves an intermediate state where the protein is simultaneously bound to two different segments [17,20,25,26], is also observed in our simulations, especially at larger values of $e_{PROT}/e_{DNA}$, although this kind of motion is not illustrated in Fig. 7.

It can easily be checked that, in contrast with 3d motion in the solvent, the number $N(t)$ of different DNA beads visited by the protein during 1d sliding very precisely follows the square root law which is expected for a random walk. For example, the solid line in Fig. 8 shows the evolution of $N(t)$ for the system with 2000 DNA beads and $e_{PROT}/e_{DNA} = 1$, obtained by averaging over 43 sliding events, which lasted more than 1 μs and during which the protein neither detached from the DNA segment for more than 0.07 µs nor reached one of the extremities of the segment. It can be seen that this solid curve very closely follows the dot-dashed line, which represents the evolution of $N(t) = \sqrt{6D_{1d}\, t}$ with a diffusion coefficient $D_{1d} = 6.7$ beads$^2$ µs$^{-1}$.

Examination of Fig. 4 indicates that the portion of time $\rho_{1d}$, during which the protein is attached to a DNA segment and experiences 1d sliding, is a monotonically increasing function of the charge protein $e_{PROT}$. In contrast, the number $N(t)$ of different DNA beads visited by the protein after a certain amount of time $t$ is *not* a monotonic function of $e_{PROT}$, and therefore of $\rho_{1d}$, as can be checked in Figs. 9 and 10. These figures display the evolution of $N(t)$ for the repulsive interaction potential of Fig. 1 and seven values of $e_{PROT}/e_{DNA}$ ranging from 0.3 to 5. In Fig 9, it is assumed that the protein is attached to bead $k$ of DNA segment $j$ if $\left\| \mathbf{r}_{j,k} - \mathbf{r}_{PROT} \right\| \le \sigma$, while the corresponding criterion is $\left\| \mathbf{r}_{j,k} - \mathbf{r}_{PROT} \right\| \le 1.5\,\sigma$ in Fig. 10. It is seen in both figures that $N(t)$ increases up to $e_{PROT}/e_{DNA} \approx 1$, then remains nearly constant up to $e_{PROT}/e_{DNA} \approx 3$, before decreasing again. The reason for this sharp decrease at large values of $e_{PROT}/e_{DNA}$ can be understood from the inspection of Fig. 11, which shows the average number of DNA beads that are simultaneously attached to the protein when it is not moving freely in solution. One observes that the number of DNA beads within $1.5\,\sigma$ of the protein is close to 2 for values of $e_{PROT}/e_{DNA}$ smaller or close to 1, which indicates that the protein forms a triangle with two successive DNA beads belonging to the same segment and separated by about $a_{DNA} = 1.78$ nm. The number of DNA beads within $1.5\,\sigma$ of the protein increases however rapidly for larger values of $e_{PROT}/e_{DNA}$, because the charge of the protein bead is sufficient to attract several DNA segments, which form a cage around it. The protein visits the DNA beads forming the cage in a short amount of time, but the slope of $N(t)$ then decreases as the protein experiences difficulties to escape the cage and visit other segments. This cage effect is strong enough for the $N(t)$ curve for $e_{PROT}/e_{DNA} = 5$ to be lower than that for the repulsive potential when the $\left\| \mathbf{r}_{j,k} - \mathbf{r}_{PROT} \right\| \leq 1.5\sigma$ criterion is considered (see Fig. 10). We will come back to this cage effect later.

Figs. 9 and 10 unambiguously show that the model exhibits facilitated diffusion, that is, the combination of 1d sliding and 3d motion leads, in a certain range of the $e_{PROT}/e_{DNA}$ ratio, to faster DNA sampling than pure 3d motion. We now assume that nature selects the fastest process and focus on the properties of the system with $e_{PROT}/e_{DNA} = 1$ (see Fig. 5 for the comparison, on a time scale much longer than in Figs. 9 and 10, of the evolution of $N(t)$ for the repulsive $V_{DNA/PROT}$ potential of Fig. 1 and the interaction potential with $e_{PROT}/e_{DNA} = 1$). Fig. 12 shows the time evolution of $N(t)$ for systems with $e_{PROT}/e_{DNA} = 1$ and increasing numbers of DNA beads, namely $mn = 990$, 2000 and 4000. As expected, the three curves coincide at short times, that is, when $N(t) << mn$. Each curve then successively displays saturation as $N(t)$ approaches $mn$. All these curves however follow the law of Eq. (3.1) with the same rate $\kappa = 1.84 \ \mu s^{-1}$, as can be checked in Fig. 13. This is rather interesting since it indicates that the observed behavior is independent of the size of the cell and can reasonably be extrapolated to larger cell sizes.

## 4 – DISCUSSION AND CONCLUSION

We proposed a dynamical model for non-specific DNA-protein interaction, which reproduces some of the observed properties of real systems and some of the hypotheses and predictions of kinetic models : (i) DNA sampling proceeds via a succession of 3d motion in the solvent, 1d sliding along the DNA sequence, short hops between neighboring sites and intersegmental transfers; (ii) facilitated diffusion takes place in a certain range of values of the protein effective charge, that is, the combination of 1d sliding and 3d motion leads to faster DNA sampling than pure 3d motion; (iii) for reasonable values of the protein effective charge, the number of base pairs visited during a single sliding event (from a few to about 20 beads, that is, from a few tens to a few hundreds base pairs) is comparable to the values deduced from single-molecule experiments [26-28,53].

The proposed model however leads to a 1d diffusion coefficient, which is too large compared to experimental values. For $e_{\text{PROT}} / e_{\text{DNA}} = 1$, we indeed obtained $D_{1d} = 6.7$ beads$^2$ $\mu$s$^{-1}$ (see Fig. 8), while experimental values are close to 5 (base pairs)$^2$ $\mu$s$^{-1}$ [27,29]. Since one bead represents 15 base pairs, this implies that the model predicts a velocity for 1d sliding, which is about one order of magnitude too large. This may be due either to the fact that real protein sliding is necessarily accompanied by geometrical rearrangements of the DNA sequence, a point which is completely neglected in the model, or to the fact that, in addition to the $E_e^{(P)}$ electrostatic interaction, the protein and the DNA sequence interact through several hydrogen bonds when the protein is sufficiently close to the sequence. This point is crucial for *specific* DNA-protein interaction (that is, target recognition) [30-35] but is again completely neglected in the proposed model for *non-specific* DNA-protein interaction.

The proposed model moreover leads to predictions, which differ from the hypotheses and conclusions of kinetic models with respect to two points. First, we observed that the number $N(t)$ of different DNA beads visited by the protein in the absence of attractive terms in $V_{\text{DNA/PROT}}$, that is when 1d sliding cannot take place, does not follow the square root law which would be expected for a purely diffusive process and is implicit in kinetic models. Fig. 6 shows that $N(t)$ instead increases linearly with rate $\kappa = 1.09$ $\mu$s$^{-1}$ till it approaches the total number of DNA beads in the cell. In Sect. 3 we tentatively ascribed this linear time dependence to either the not-so-infrequent collisions with DNA or to the fact that our dynamical model takes electrostatic interactions into account, while kinetic models usually do not. Moreover, kinetic models predict that facilitated diffusion might speed up the search time by a factor of approximately 30 [20], while we obtained a maximum factor of about 2 ($\kappa = 1.84$ $\mu s^{-1}$ for $e_{PROT}/e_{DNA} = 1$ against $\kappa = 1.09$ $\mu s^{-1}$ for the repulsive potential). This discrepancy with kinetic models might result from the above-mentioned linear dependence of $N(t)$ in the absence of 1d sliding, which implies that 3d motion is as not as inefficient compared to 1d sliding as in kinetic models.

The proposed model can (must) be improved with respect to several points. To our mind, the roughest approximation concerns the protein, which we describe as a single bead with an electric charge $e_{PROT}$ placed at its center. For large values of $e_{PROT}$, this leads to the cage effect discussed in Sect. 3 (see Fig. 11) and to too frequent intersegmental transfers. Without trying to provide as detailed a description as, for example, in [38,39,54], a better approximation would still consist in considering the protein as a set of interconnected beads with a certain charge distribution. It will be interesting to check whether the rates $\kappa$ and the maximum search time speed up factor obtained with the improved model match those of the present one. Moreover, the present model describes how certain proteins like transcriptions factors proceed via a succession of 1d sliding and 3d motion to sample the DNA sequence, but it provides no clue to how the protein recognizes and fixes to its specific target during 1d sliding. Incorporating this point in the model will certainly require a finer description of the DNA sequence : a bead will no longer describe a set of 15 successive base pairs, but rather a single base pair, and each DNA bead will interact with the protein beads via heterogeneous distributions of charges and hydrogen bonds. Other possible improvements include consideration of the torsion of the DNA sequence, introduction of some interaction between DNA transient bubbles and 1d sliding of the protein, *etc*…

**FIGURE CAPTIONS**

**Figure 1** (color online) : Plot, as a function of the distance $\left\| \mathbf{r}_{j,k} - \mathbf{r}_{\mathrm{PROT}} \right\|$ between the two beads, of the interaction potential $V_{\mathrm{DNA/PROT}}$ between the protein bead and bead $k$ of DNA segment $j$, for three different values of $e_{\mathrm{PROT}} / e_{\mathrm{DNA}}$ (0.3, 2 and 5) and a purely repulsive potential, which is just the repulsive part of the potential with $e_{\mathrm{PROT}} / e_{\mathrm{DNA}} = 0.3$. $V_{\mathrm{DNA/PROT}}$ is expressed in eV and $\left\| \mathbf{r}_{j,k} - \mathbf{r}_{\mathrm{PROT}} \right\|$ in nm. Note that the three curves with $e_{\mathrm{PROT}} / e_{\mathrm{DNA}} = 0.3$, 2 and 5 all display a minimum located at $\left\| \mathbf{r}_{j,k} - \mathbf{r}_{\mathrm{PROT}} \right\| = 5.04$ nm, close to $\sigma = a_{\mathrm{DNA}} + a_{\mathrm{PROT}} = 5.28$ nm.

**Figure 2** (color online) : Profile of the number of DNA beads per unit volume as a function of the distance $r$ from the centre of the cell after an integration time of 30 µs. The maximum of the curve was arbitrarily scaled to 1. This profile was averaged over 64 different trajectories with 2000 DNA beads.

**Figure 3** (color online) : Comparison of results obtained with different time steps $\Delta t$. Both plots show the evolution of $N(t)$, the number of different DNA beads visited by the protein at time $t$. It is considered that a DNA bead and the protein are in contact if the distance between the centers of the two beads is smaller than $\sigma = a_{\mathrm{DNA}} + a_{\mathrm{PROT}} = 5.28$ nm. The top plot shows the evolution of $N(t)$ for the system with 2000 DNA beads, $e_{\mathrm{PROT}} / e_{\mathrm{DNA}} = 1$ and time steps $\Delta t = 25$ and 100 ps. The bottom plot shows the evolution of $N(t)$ for the system with 4000 DNA beads, $e_{\mathrm{PROT}} / e_{\mathrm{DNA}} = 1$ and time steps $\Delta t = 100$ and 400 ps. Each curve was averaged over 6 different trajectories.

**Figure 4** (color online) : Plot, as a function of the ratio $e_{\mathrm{PROT}} / e_{\mathrm{DNA}}$, of the portion of time $\rho_{\mathrm{1d}}$ during which the protein remains attached to a DNA bead. The abscissa axis actually corresponds to the variation of $e_{\mathrm{PROT}}$ at constant $e_{\mathrm{DNA}}$. Circles and lozenges denote results obtained with, respectively, the $\left\| \mathbf{r}_{j,k} - \mathbf{r}_{\mathrm{PROT}} \right\| \leq \sigma$ and $\left\| \mathbf{r}_{j,k} - \mathbf{r}_{\mathrm{PROT}} \right\| \leq 1.5\,\sigma$ criterions for interacting beads. The point at $e_{\mathrm{PROT}} / e_{\mathrm{DNA}} = 0$ was obtained with the repulsive potential of

Fig. 1. Each point was averaged over 12 different trajectories propagated for 100 μs for the system with 2000 beads.

**Figure 5** (color online) : Evolution of $N(t)$, the number of different DNA beads visited by the protein at time $t$, for the system with 2000 DNA beads and the interaction potential $V_{\text{DNA/PROT}}$ with $e_{\text{PROT}}/e_{\text{DNA}} = 1$ (solid line) and the repulsive potential of Fig. 1 (dotted line). It was assumed that the protein is attached to bead $k$ of DNA segment $j$ if $\left\| \mathbf{r}_{j,k} - \mathbf{r}_{\text{PROT}} \right\| \leq \sigma$. Each curve was averaged over 6 different trajectories. The time evolution of a diffusive process with $D = 100$ beads$^2$ μs$^{-1}$ is also shown for the sake of comparison (dash-dotted line).

**Figure 6** (color online) : Plot of $\ln\left(1 - N(t)/(m\,n)\right)$ as a function of $t/(m\,n)$ for the system with $m\,n = 2000$ DNA beads and the repulsive potential of Fig. 1. $N(t)$ corresponds to the dotted line in Fig. 5. The dot-dashed straight line represents the same plot for the expression of $N(t)$ in Eq. (3.1) and a rate $\kappa = 1.09$ μs$^{-1}$.

**Figure 7** (color online) : Typical protein trajectory for the system with 2000 DNA beads and the ratio $e_{\text{PROT}}/e_{\text{DNA}} = 1$. This plot indicates, at each time, to which bead of which DNA segment the protein is eventually attached. Time intervals for which no position is indicated correspond to those periods where the protein is moving in the solvent. It was assumed that the protein is attached to bead $k$ of DNA segment $j$ if $\left\| \mathbf{r}_{j,k} - \mathbf{r}_{\text{PROT}} \right\| \leq \sigma$.

**Figure 8** (color online) : Evolution of the number $N(t)$ of different DNA beads visited by the protein during 1d sliding. Calculations were performed with 2000 DNA beads and the ratio $e_{\text{PROT}}/e_{\text{DNA}} = 1$. $N(t)$ was averaged over 43 sliding events with the following properties : (i) each sliding event lasted more than 1 μs, (ii) the protein did not separate from the DNA segment by more than $\sigma$ during more than 0.07 μs, (iii) the protein bead did not reach one of the extremities of the DNA segment. The dot-dashed line corresponds to a diffusion coefficient $D_{1d} = 6.7$ beads$^2$ μs$^{-1}$.

**Figure 9** (color online) : Evolution of the number $N(t)$ of different DNA beads visited by the protein, for seven values of $e_{\text{PROT}}/e_{\text{DNA}}$ ranging from 0.3 to 5 and for the repulsive

DNA/protein interaction potential of Fig. 1. Each curve was averaged over 12 different trajectories for the system with 2000 beads. It was assumed that the protein is attached to bead $k$ of DNA segment $j$ if $\left\| \mathbf{r}_{j,k} - \mathbf{r}_{\text{PROT}} \right\| \leq \sigma$.

**Figure 10** (color online) : Same as Fig. 9, except that it is considered that the protein is attached to bead $k$ of DNA segment $j$ if $\left\| \mathbf{r}_{j,k} - \mathbf{r}_{\text{PROT}} \right\| \leq 1.5\,\sigma$ instead of $\left\| \mathbf{r}_{j,k} - \mathbf{r}_{\text{PROT}} \right\| \leq \sigma$.

**Figure 11** (color online) : Plot, as a function of the ratio $e_{\text{PROT}} / e_{\text{DNA}}$, of the average number of DNA beads that are attached to the protein when it does not move freely in solution. The abscissa axis actually corresponds to the variation of $e_{\text{PROT}}$ at constant $e_{\text{DNA}}$. Circles and lozenges denote results obtained with, respectively, the $\left\| \mathbf{r}_{j,k} - \mathbf{r}_{\text{PROT}} \right\| \leq \sigma$ and $\left\| \mathbf{r}_{j,k} - \mathbf{r}_{\text{PROT}} \right\| \leq 1.5\,\sigma$ criterions for interacting beads. The point at $e_{\text{PROT}} / e_{\text{DNA}} = 0$ was obtained with the repulsive potential of Fig. 1. Each point was averaged over 12 different trajectories propagated for 100 µs for the system with 2000 beads.

**Figure 12** (color online) : Evolution of $N(t)$, the number of different DNA beads visited by the protein at time $t$, for the system with $e_{\text{PROT}} / e_{\text{DNA}} = 1$ and 990, 2000 and 4000 DNA beads. It was assumed that the protein is attached to bead $k$ of DNA segment $j$ if $\left\| \mathbf{r}_{j,k} - \mathbf{r}_{\text{PROT}} \right\| \leq \sigma$. Each curve was averaged over 6 different trajectories.

**Figure 13** (color online) : Solid line : plot of $\ln\left(1 - N(t)/(mn)\right)$ as a function of $t/(mn)$ for the system with $e_{\text{PROT}} / e_{\text{DNA}} = 1$ and 990, 2000 and 4000 DNA beads (the curves for 2000 and 4000 beads nearly superpose). $N(t)$ corresponds to the curves in Fig. 12. The dot-dashed straight line represents the same plot for the expression of $N(t)$ in Eq. (3.1) and a rate $\kappa = 1.84 \ \mu\text{s}^{-1}$.

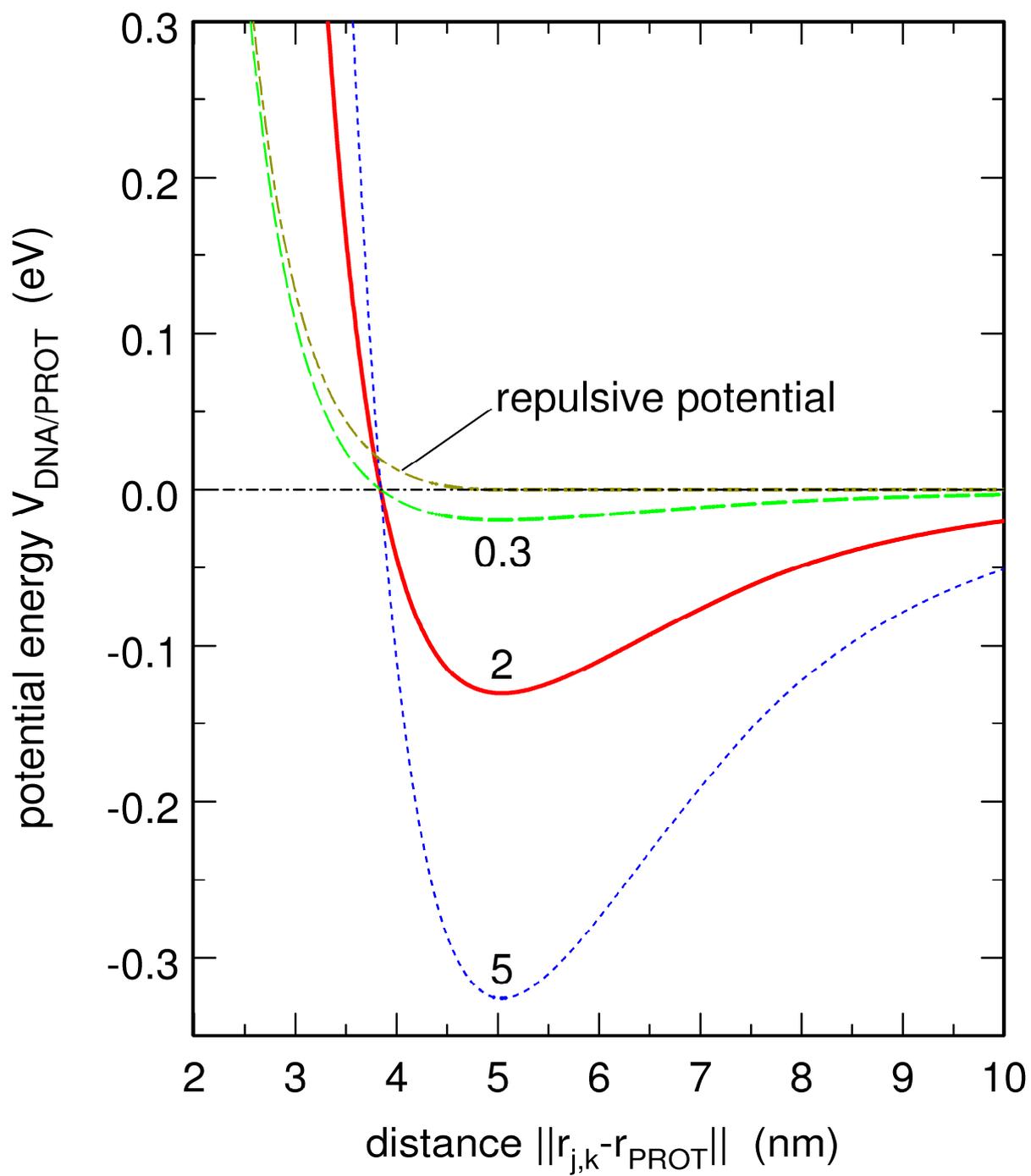

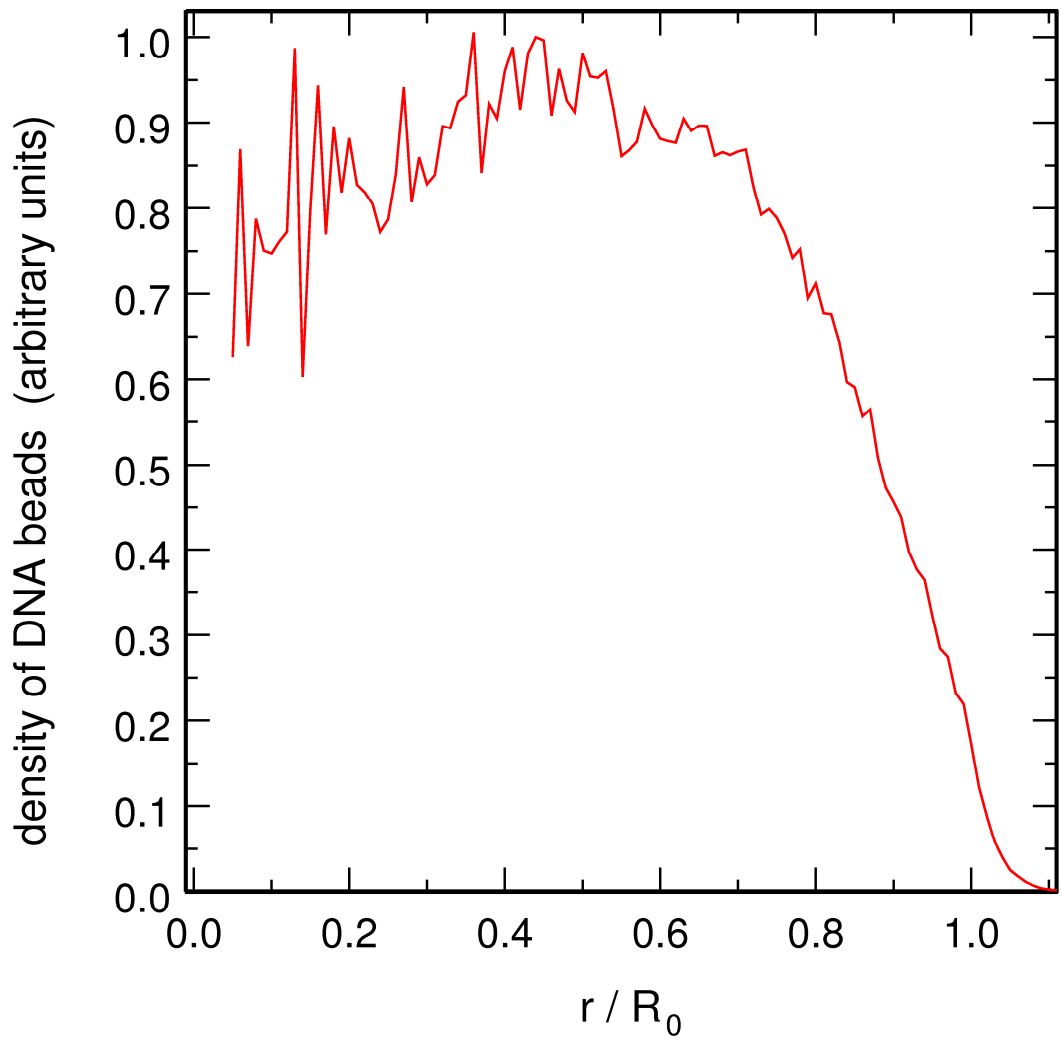

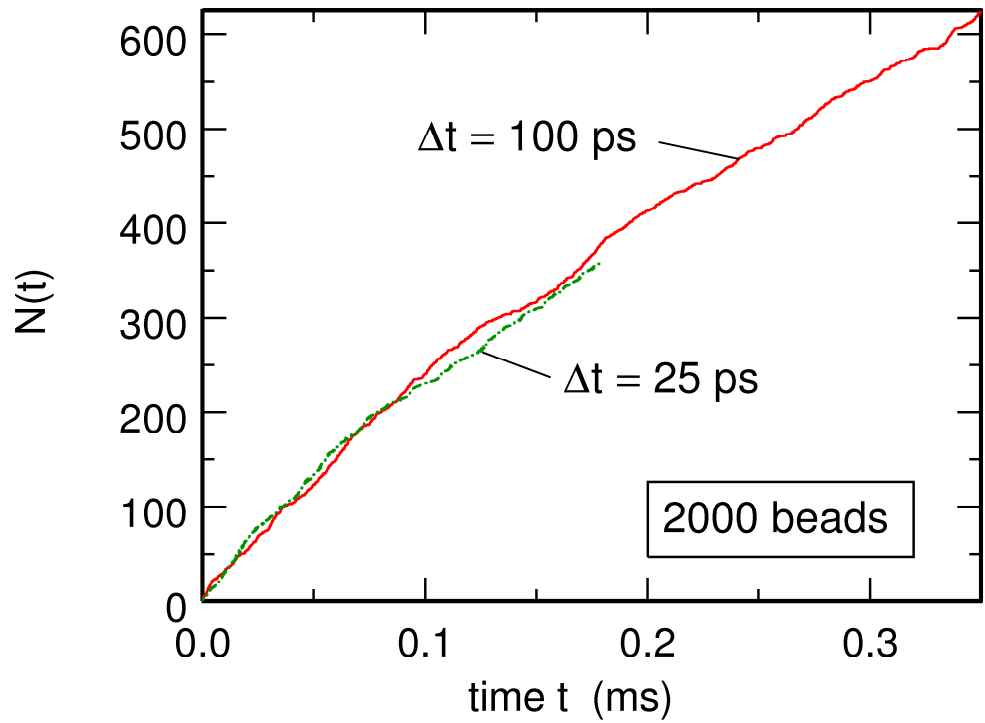

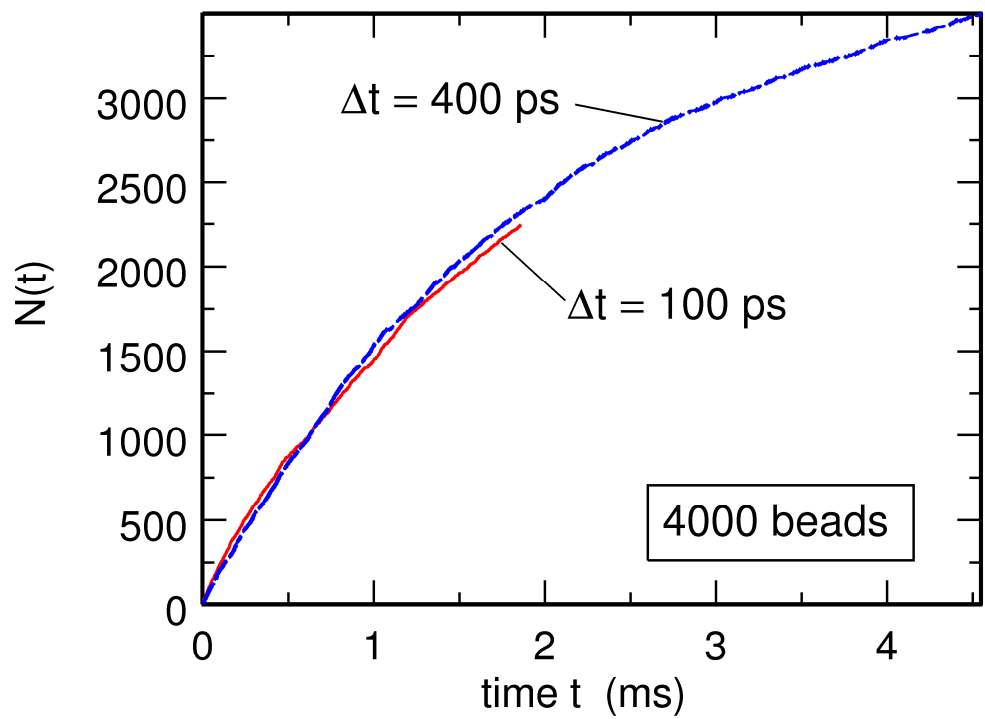

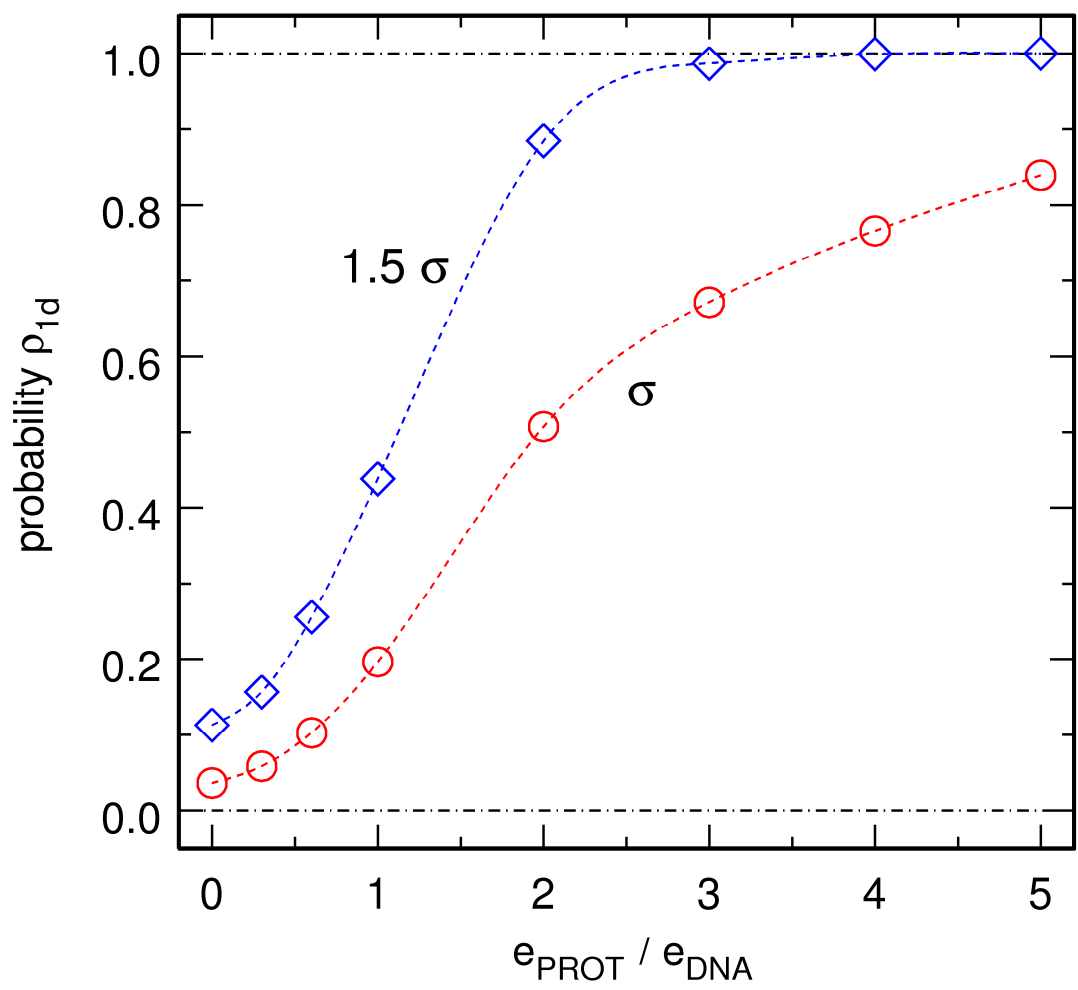

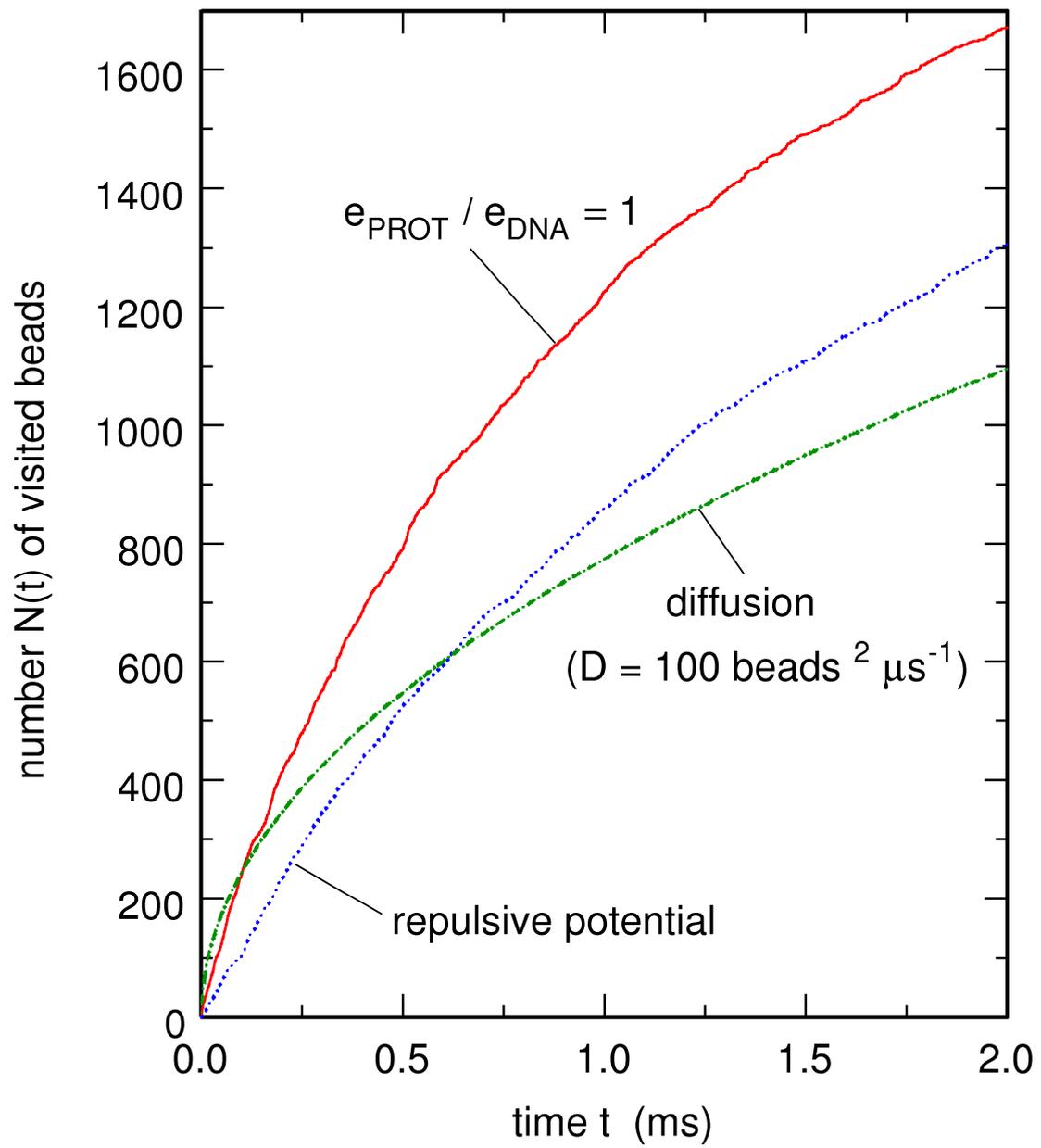

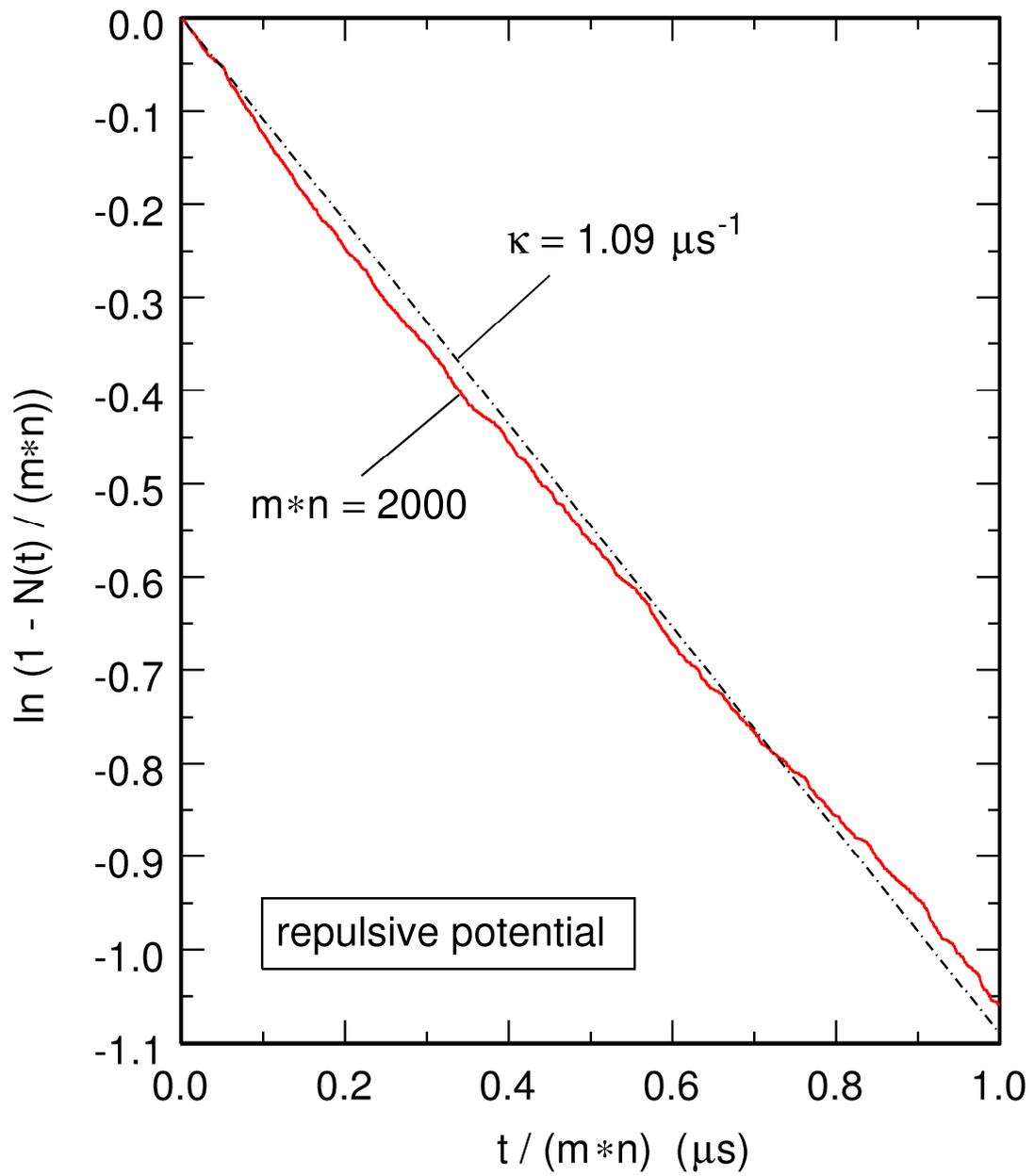

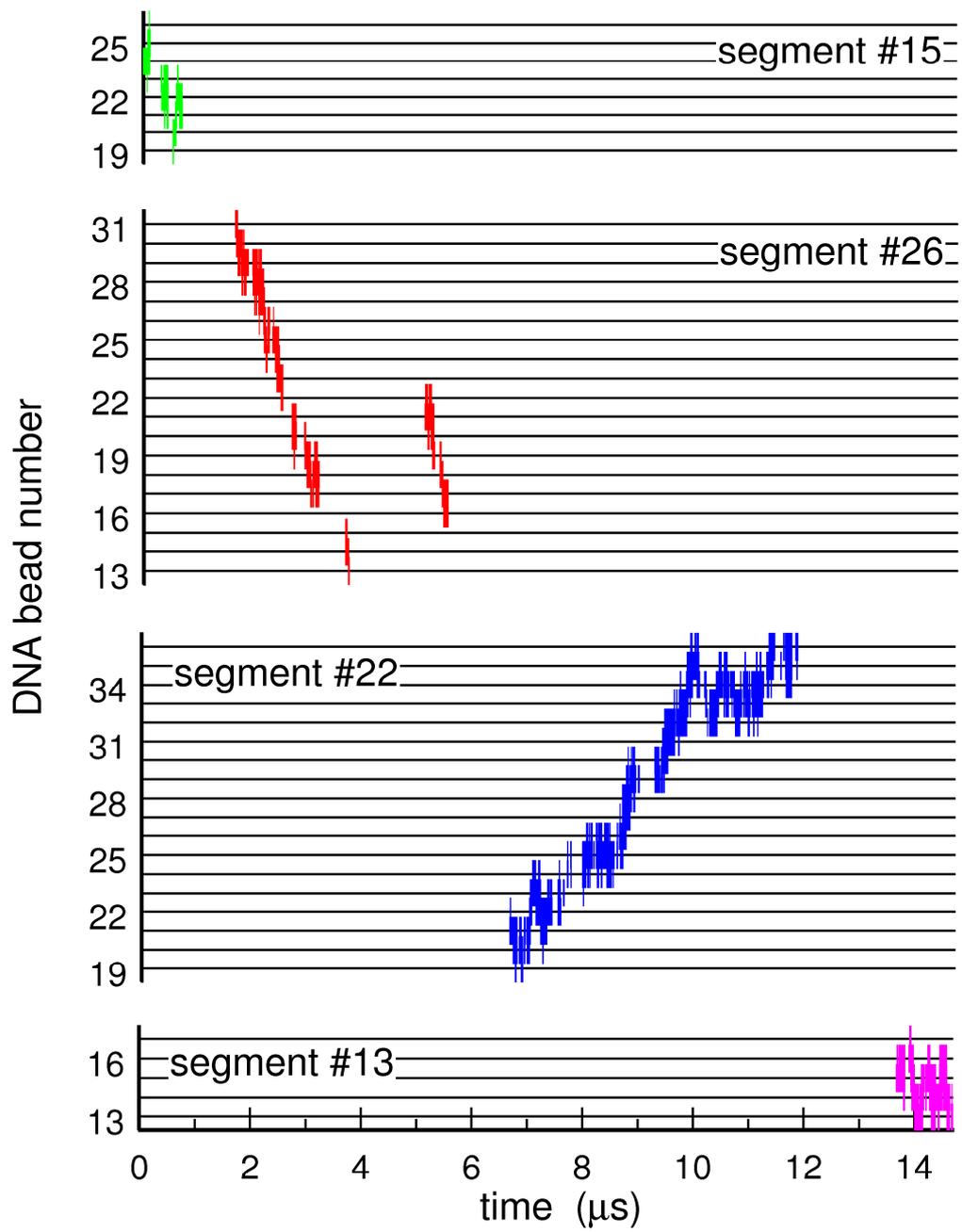

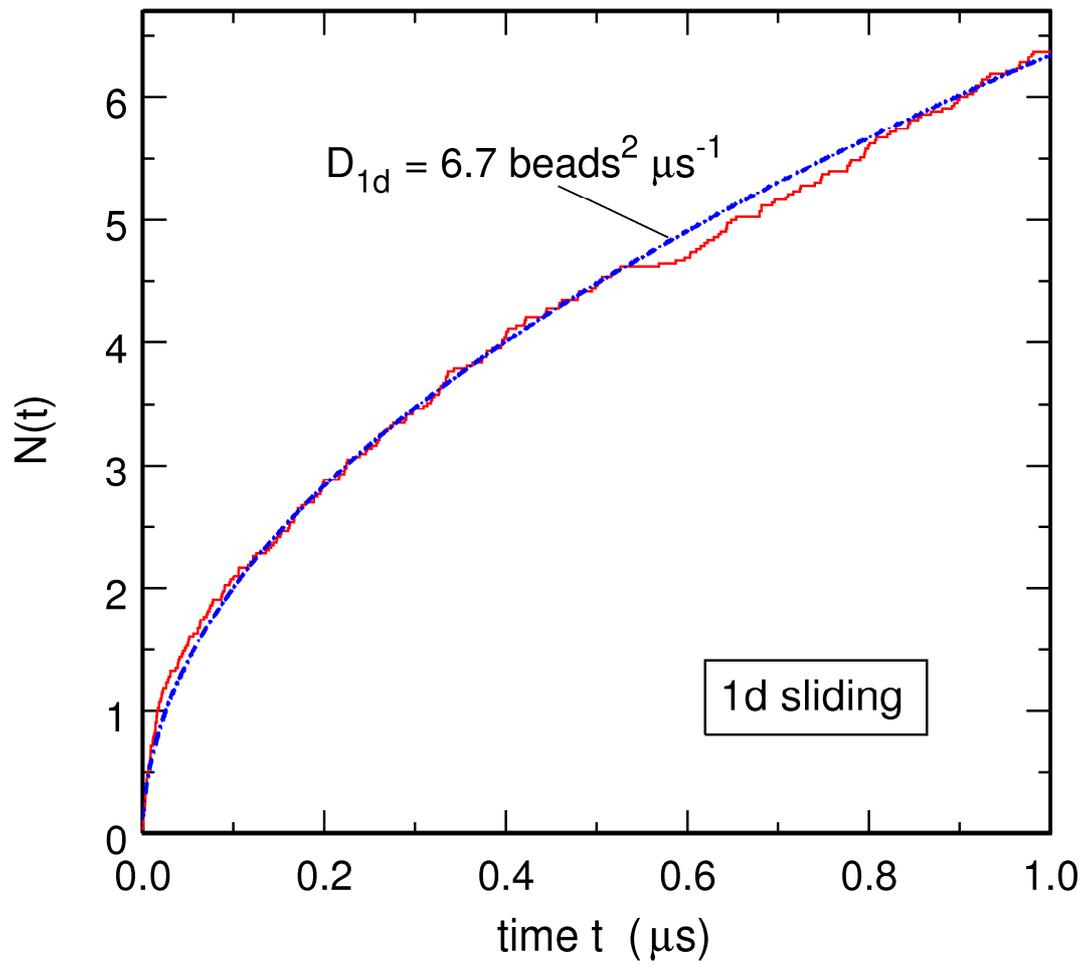

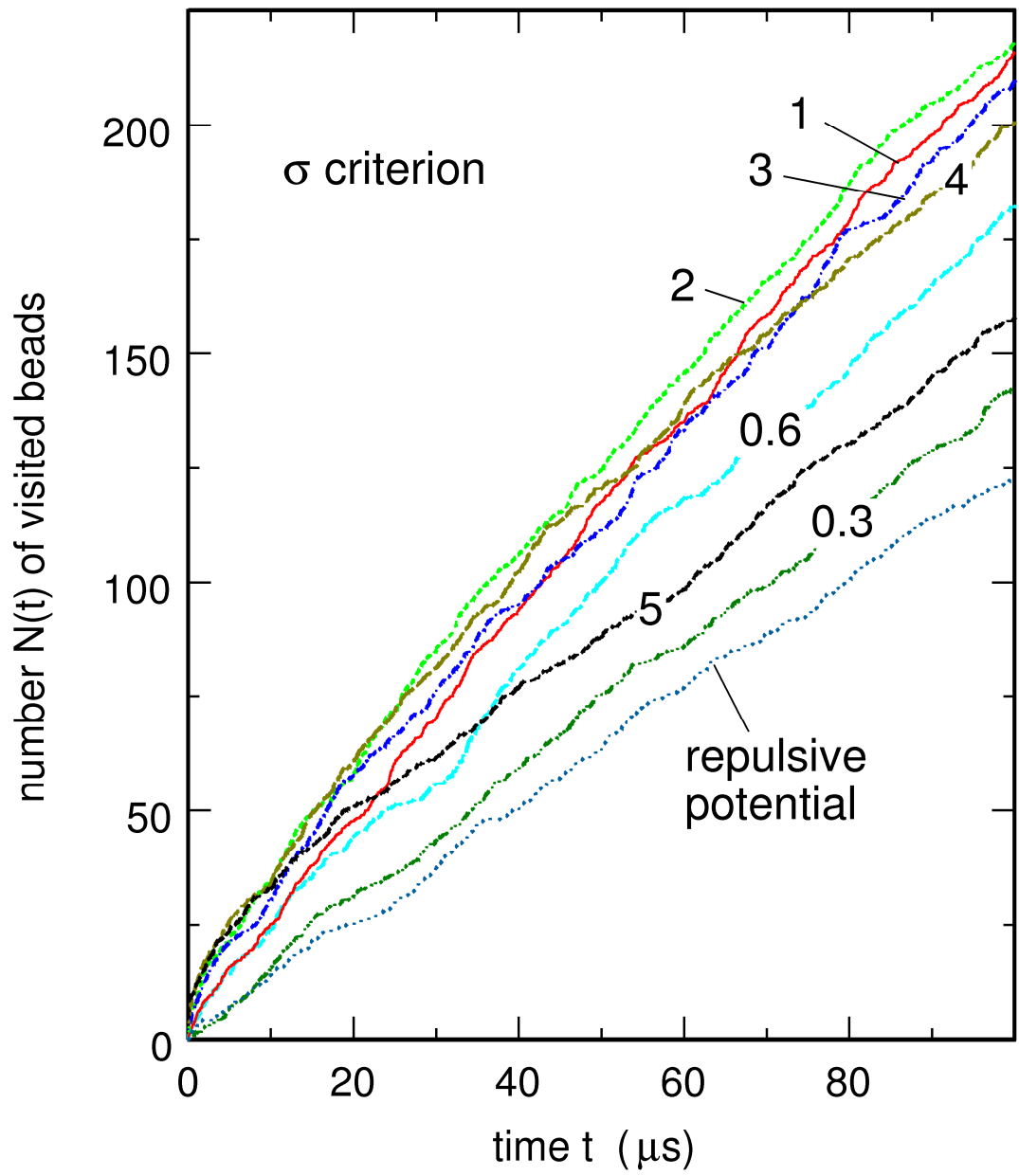

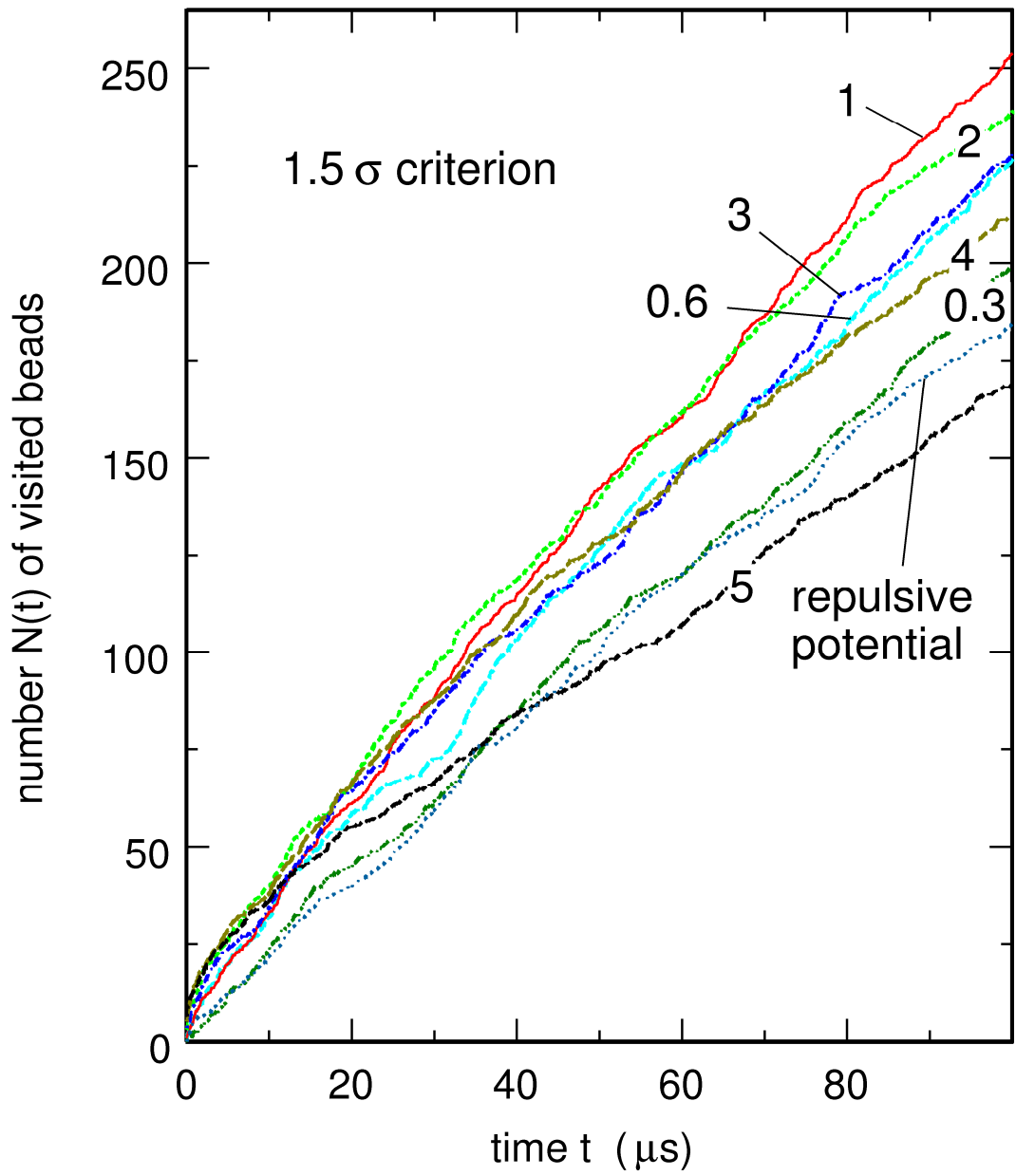

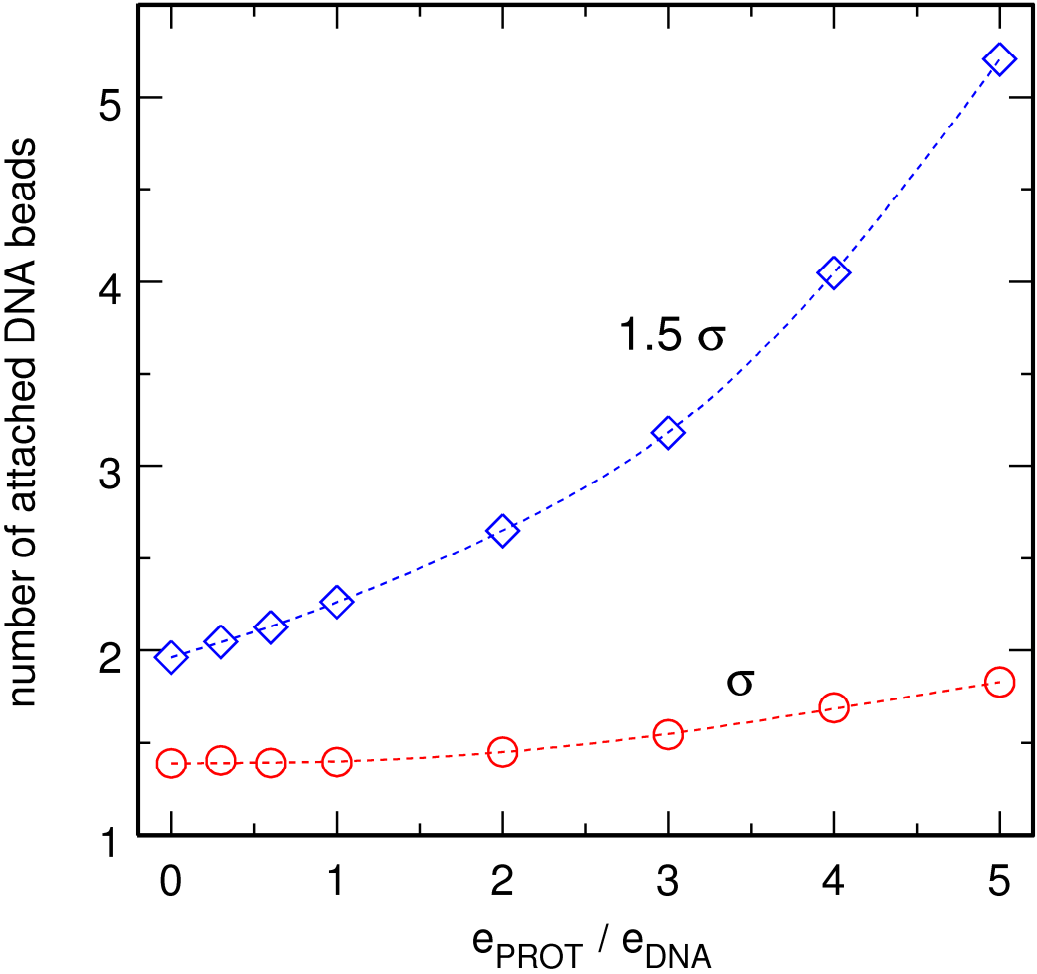

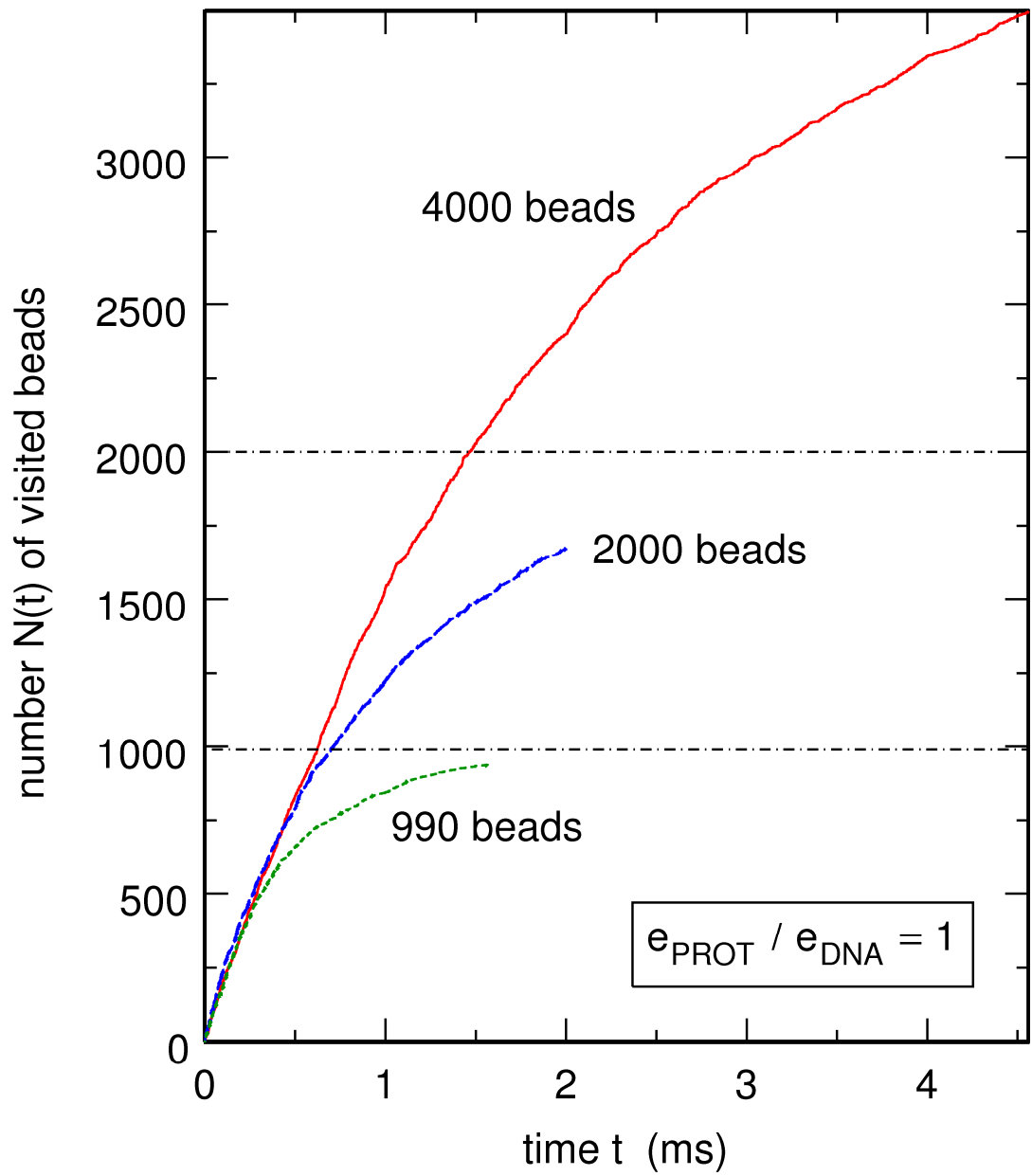

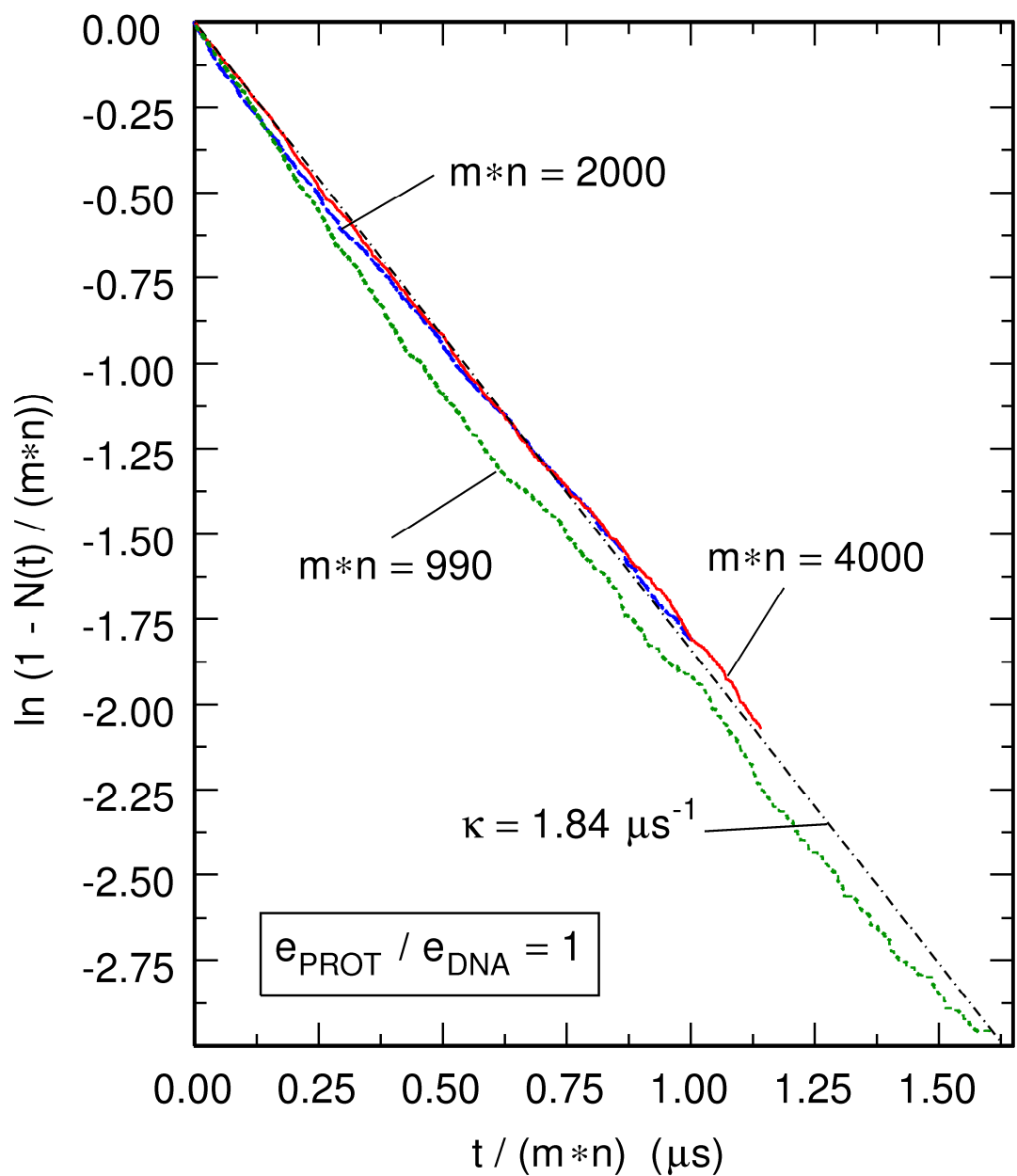